\documentclass{elsart}

\usepackage{graphicx,natbib,amssymb}

\journal{New Astronomy}
\begin{document}

\begin{frontmatter}

\title{Spectroscopy of the hot pulsating star $\beta$\,Cephei.\\ Velocities and EWs from C,N,O and Si lines\thanksref{osser}}

\thanks[osser]{Based on observations collected at the 91~cm telescope of the INAF - Osservatorio Astrofisico di Catania}

\author[oact]{G. Catanzaro\corauthref{cor1}},
\corauth[cor1]{Corresponding author.}
\ead{gca@oact.inaf.it}
\author[unict]{F. Leone},
\author[oact]{I. Bus\'a} and
\author[oact]{P. Romano}

\address[oact]{INAF-Osservatorio Astrofisico di Catania, via S. Sofia 78, 95123 Catania, Italy}
\address[unict]{Sezione Astrofisica-Dipartimento di Fisica e Astronomia, Universit\`a di Catania,
             via S. Sofia 78, 95123 Catania, Italy}

\begin{abstract}
   Frequencies in oscillating $\beta$ Cephei stars are usually inferred by means of 
   radial velocities measured from the Si{\sc iii} triplet 
   $\lambda \lambda$\,4552\,-\,4574 {\AA}. These lines, relatively insensitive to the variation 
   of T$_{\rm eff}$ through a pulsation cycle, show small equivalent width variations.

   In this study we aimed to verify if the behavior of radial velocities and equivalent 
   widths measured from other ions are compatible with the one observed from Si{\sc iii} 
   lines and than to verify the possible vertical stratification along the stellar atmosphere.
   For this reason we selected from our spectra  a number of, unblended and well isolated, 
   C{\sc ii}, N{\sc ii} and O{\sc ii} lines besides the famous Si{\sc iii} triplet. All those
   lines cover the range in optical depth between -2.1 and -0.5.

   Unfortunately, we did not find any differences in the radial velocities behavior line-by-line
   and then we derived the frequency of the principal radial mode combining all the 
   velocities derived from each spectral line separately. The inferred frequency was
   $f_1$\,=\,5.249677\,$\pm$\,0.000007~c~d$^{-1}$.   

   Another important task we would like to accomplish with this paper is to make available to the 
   community our large sample of spectroscopic data, that is 932 velocities and equivalent widths 
   measured from our sample of C, N, O and Si lines. All the spectra were acquired at the  
   1-meter class telescope of the stellar station of the {\it INAF - Osservatorio 
   Astrofisico di Catania}, in the period starting from July, the 27$^{th}$ 2005 to November, the 
   1$^{st}$ 2006.
\end{abstract}


\end{frontmatter}

\section{Introduction}
The star $\beta$ Cephei (HD\,205021) is the prototype of a class of hot
pulsating variables. For complete reviews on $\beta$ Cephei stars, we refer
the readers to \citet{aerts03}, who describe in detail
the line-profile variations and to \citet{sterken92},
who give an overview of the photometric behavior of all the $\beta$ Cep stars
known up to that time. 

From the spectroscopic point of view, the data used to recover frequencies of
oscillations are commonly the radial velocities. Checking through the most recent 
literature published on this topic, it is easy to realize that the common guideline is 
to derive the velocities from the lines of the Si{\sc iii} triplet at 
4552, 4567 and 4574 {\AA}. There are three common reasons for this choice: they 
are strong lines not affected very much by blending and almost insensitive to the 
temperature variations \citep{deridder02}. 

Up to the study by \citet{aerts94}, $\beta$ Cephei has been believed 
to oscillate with only a single radial mode. 
In their analysis these authors revealed the multi-periodicity
of this star detecting three frequencies in total: $f_1$\,=\,5.2497104~c~d$^{-1}$, 
$f_2$\,=\,5.385~c~d$^{-1}$ and $f_3$\,=\,4.920~c~d$^{-1}$. Later on, \citet{telting97} 
by a CLEANing analysis of the same data added two more 
frequencies: $f_4$\,=\,5.083~c~d$^{-1}$ and $f_5$\,=\,5.417~c~d$^{-1}$.

Further, the star $\beta$ Cep (V\,=\,3.2) is actually a complicated multiple system. 
In fact the star, beside to have a visual companion (V\,=\,7.9) at a distance of 13.4'', 
is also a member of a spectroscopic system whose second star was discovered by
\citet{gezari72}, using speckle interferometry, at a distance of $\approx$0.25'' 
(V\,=\,6.6). The parameters of the close binary orbit have been determined later by 
\citet{pigu92} from speckle interferometry and the variations in the 
pulsation period, due to the so-called light time effect. Recent speckle
measurements by \citet{hartkopf01} place the position of the
companion to about 0.1'' from the primary. More observation are still necessary to improve
the orbital solution.

Another question, people are still debating, is the nature of the random variable H$_\alpha$
emission which was the cause of the Be status assigned to $\beta$ Cep. The problem
was that typically Be stars are rapid rotators while $\beta$ Cep has an equatorial
velocity of 26~km~s$^{-1}$ \citep{morel06}. \citet{hadrava96},
attempted to solve this problem argued that the observed emission is due to the 
secondary component of the spectroscopic binary which is probably a Be star. They,
for the first time, separated the emission and absorption component of the H$_\alpha$
profile. Recently, \citet{schnerr06} confirmed this result
using data obtained at Nordic Optical Telescope. Unfortunately, they obtained only one
spectrum and, then, no conclusion about the variability of the emission could be drawn.

Starting from the pioneering work by \citet{osaki71}, a lot of studies have been 
undertaken with the aims to infer the oscillation modes from the analysis of line
profile variation (LPV). It is generally accepted that the weakness of these studies
is the lack in the knowledge of how some important thermodynamic quantities change during
the pulsation. For instance, during each pulsation cycle effective temperature, 
surface gravity and microturbulent velocity are expected to change with time and than 
the knowledge of their effect on the line profiles could avoid errors in the modes 
determination.

It is a matter of fact that, because of its low rotational velocity, the spectrum of 
$\beta$ Cephei is full of lines generated by various chemical elements at different
depths in the photosphere. The analysis of the radial velocity inferred from a sample
of lines opportunely selected could be useful to study the velocity stratification
along the photosphere. A similar approach has been tempted by \citet{baldry98} in a 
spectroscopy study of the roAp star $\alpha$\,Cir. In this study latter authors selected
a sample of spectral region containing various metallic lines and they found a dependence
of the amplitudes with the depth along the atmosphere.   

With the aim to infer if a similar stratification is present also in $\beta$ Cep, we selected
from our spectra as many as possible lines not affected by blends, sufficiently isolated
to allow an easy measurements and well distributed along the atmosphere.
A complete list of selected lines is reported in Tab.~\ref{res_fit}.
In this paper we present a large time-resolved dataset of line-by-line
radial velocities collected from July 2005 to November 2006, for a total of 932 spectra 
and, moreover, for the first time we present also their equivalent widths.

Velocities and equivalent widths can be retrieved from the CDS via anonymous ftp.
\begin{table}
\caption{Journal of observations. JD are in the form JD\,-\,2450000, {\it n} is
the number of spectra collected in each night.} 
\label{obs_log}     
\centering                
\begin{tabular}{c c c c}    
\hline\hline              
 JD & n &  JD & n \\
\hline                    
3579 & 56 & 3933 & 49 \\
3580 & 44 & 4013 & 35 \\
3581 & 46 & 4016 & 30 \\
3635 & 65 & 4017 & 26 \\
3672 & 64 & 4018 & 41 \\
3673 & 95 & 4019 & 43 \\
3674 & 82 & 4021 & 19 \\
3871 & 58 & 4022 & 18 \\
3872 & 61 & 4023 &  5 \\
3873 & 64 & 4041 & 31 \\
\hline      
\end{tabular}
\end{table}

\section{Observations and data reduction}
All the data analyzed in this study have been obtained at the 91 cm
telescope of the {\it INAF\,-\,Osservatorio Astrofisico di Catania}. 
The telescope is fiber linked to a REOSC echelle spectrograph, which allows to obtain
R\,=\,20\,000 spectra in the range 4300-6800 {\AA}. The resolving power
has been checked using emission lines of the Th-Ar calibration lamp.
Spectra were recorded on a thinned, back-illuminated (SITE) 
CCD with 1024 x 1024 pixels of 24~$\mu$m size, typical readout noise of
6.5 e$^{-}$ and gain of 2.5 ph/ADU. 

Our observations (932 spectra in total) were spread over a baseline greater than 
one year, precisely from July, the 27$^{th}$ 2005 (JD\,=\,2453579) to November, 
the 1$^{st}$ 2006 (JD\,=\,2454041), for a total of 462.14 days. The exposure 
time was set to 5 minutes, with the exception of a couple of nights in October 2006 
during which we had to increase it to 10 minutes because of the bad weather condition. 
This leads to a temporal resolution of $\approx$~2$\%$ and $\approx$~4$\%$, respectively, 
of the pulsational period, small enough in both cases to avoid phase smearing effects. 
The signal-to-noise ratio of our spectra resulted always above 150. 

The stellar spectra, calibrated in wavelength and with the continuum normalized
to a unity level, were obtained using standard data reduction procedures for 
spectroscopic observations within the NOAO/IRAF package, that is: bias frame 
subtraction, trimming, scattered light correction, flat-fielding, fitting traces
and orders extraction and, finally, wavelength calibration. IRAF package 
{\it rvcorrect} has been used to include the velocity correction due to the Earth's
motions, all the spectra were then reduced into the heliocentric rest of frame.

Systematic errors on radial velocities have been estimated observing stars with 
constant and well known radial velocity taken from the list of standard stars 
published by \citet{udry99}: HD\,12929, HD\,20902, HD\,186791 and 
HD\,206778. Whenever it was not possible to obtain a spectrum of a standard 
star, we evaluated the systematic shift from the wavelength positions 
of the strong telluric lines between 6275 {\AA} and 6320 {\AA}.

  \begin{figure*}
   \centering
   \hbox{
   \includegraphics[width=7cm]{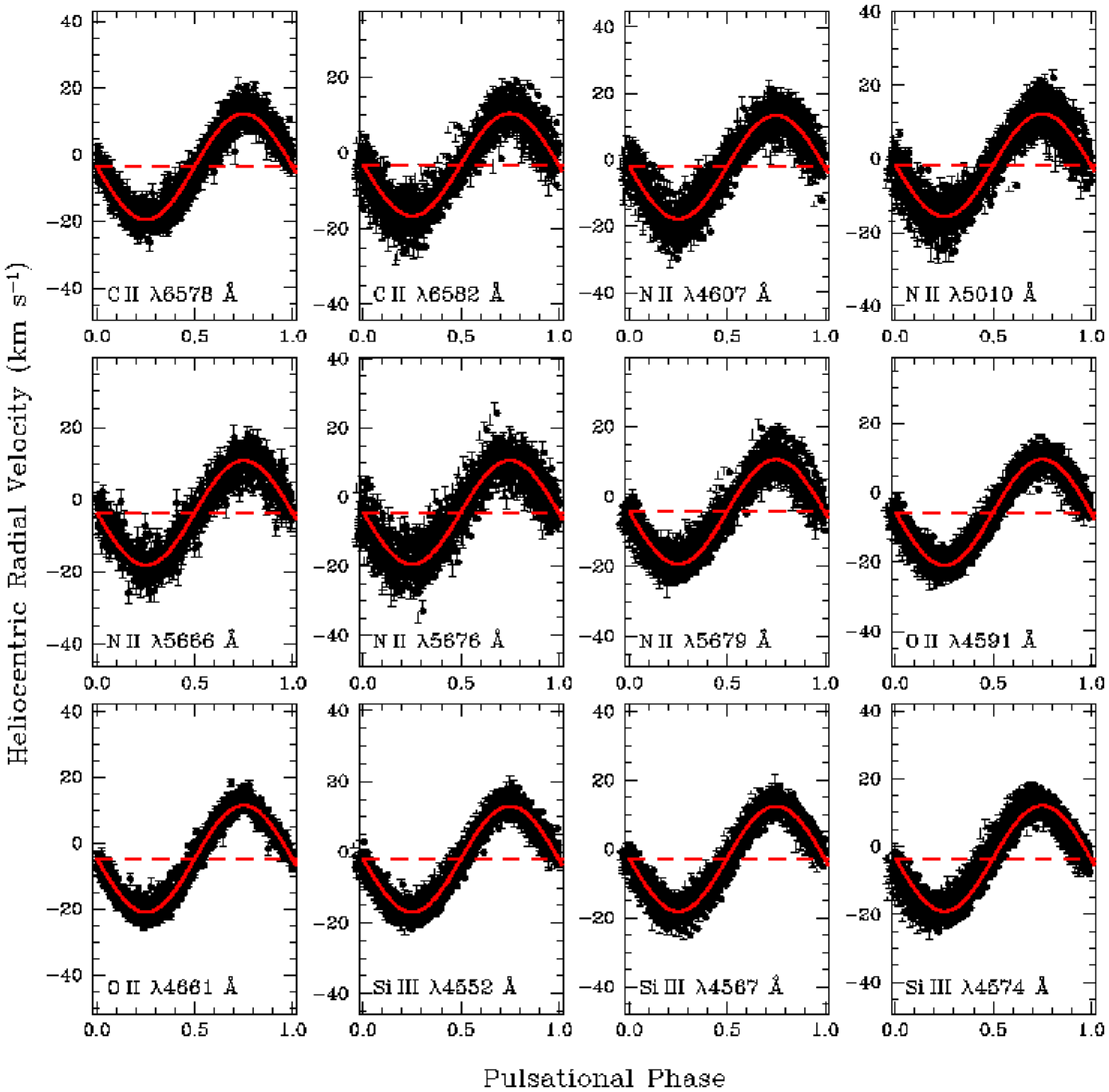}
   \includegraphics[width=7cm]{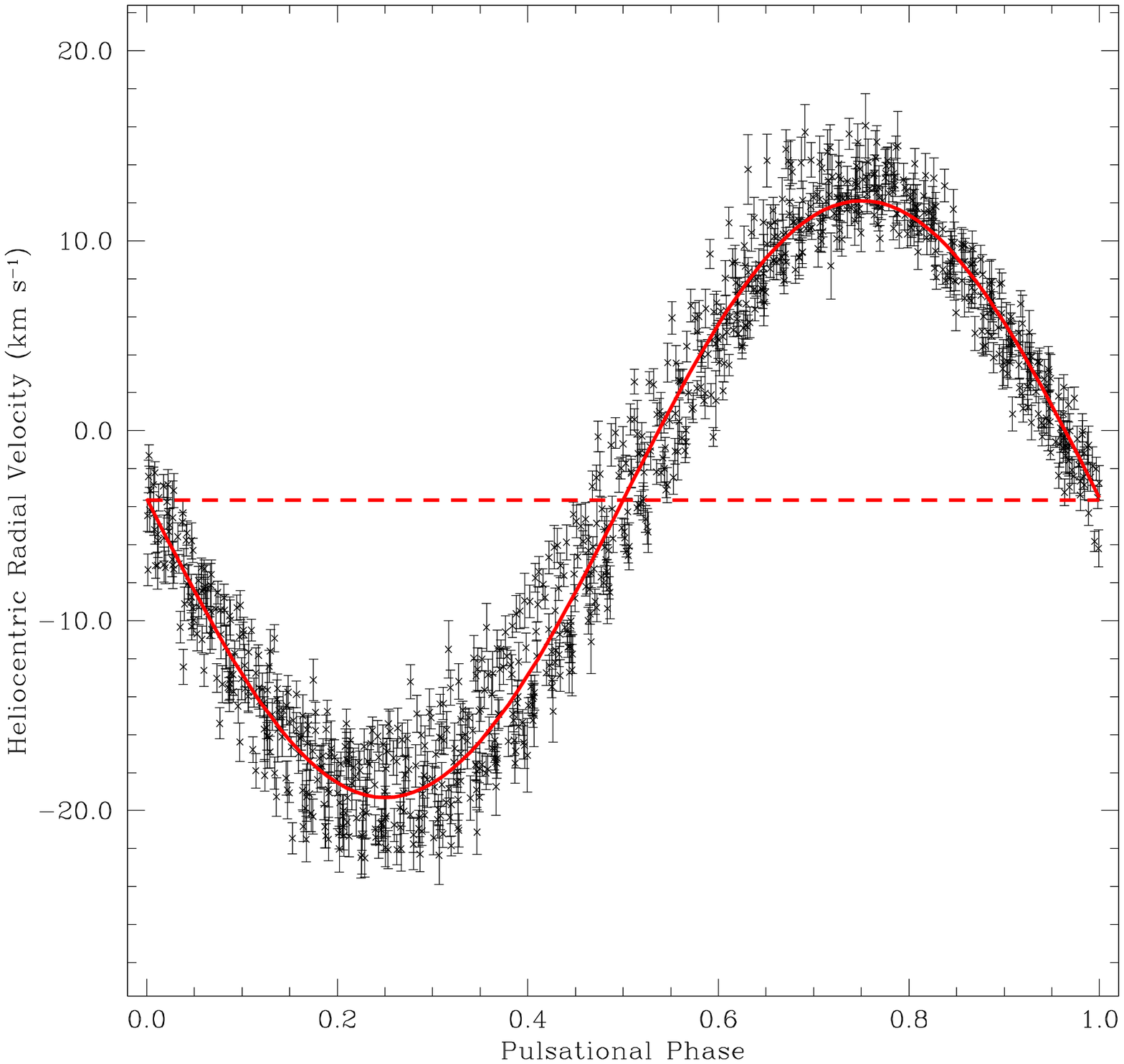}
       }
   \caption{Phase diagrams of the radial velocity variations derived by
            carbon, nitrogen, oxygen and silicon lines (left side)
            and velocities computed from average over all the lines selected
            (right side). Parameters of the over-imposed sinusoidal fits 
            are reported in Tab.~\ref{res_fit}. Error bars extend for $\pm$~1~$\sigma$.}
   \label{pdm}
   \end{figure*}
\section{Radial velocities and equivalent widths measurements}
\label{vrad_s}
As we stated before, radial velocities in this class of 
pulsating stars are usually measured from the Si{\sc iii} triplet. Here we have 
considered the heliocentric V$_{\rm rad}$ measured from lines of other ions with 
the final aim to verify if their behavior is different line-by-line. Thus, we selected all the strong,
unblended lines present in the spectral range covered by our spectra.
These lines are reported in the first column of Tab.~\ref{res_fit} and are, precisely, 
2 lines belonging to C{\sc ii}, 5 to N{\sc ii}, 2 to O{\sc ii} and 3 lines to 
Si{\sc iii}.

For each of these lines we adopted as radial velocity the value obtained
from the barycentric wavelength which has been converted in radial velocity
via the classical Doppler shift formula. The integrations evaluated to compute
$\lambda_{\rm bar}$ and the equivalent width (EW) of each line extend over a 
carefully chosen, constant region in wavelength and they have been evaluated
after a local re-normalization of the continuum close to the edges of the line.
This procedure let us to minimize the systematic effects due to the arbitrary
choice of the lines boundaries or the position of the continuum. 
Errors on V$_{\rm rad}$ and EW  have been computed from a formal application of 
the error propagation rules to the used equations, assuming that the main source of error is 
the noise in the observed spectrum. 

Single values of radial velocities and equivalent widths are presented only 
in the online tables\footnote{Table 1 reports lines from C{\sc ii} 6578 to 
N{\sc ii} 5676, Table 2 reports lines from N{\sc ii} 5679 to Si{\sc iii} 4574},
available at the {\it CDS}, which are organized as follow: the first column
reports the heliocentric julian date, the others reported line-by-line the
measured radial velocities and equivalent widths together with their errors. 
Blank space means a discarded line profile contaminated by cosmic rays.

\begin{table*}
\tiny
\caption{Parameters and relative errors of the sinusoidal fits presented in Fig.~\ref{pdm} and in Fig.~\ref{var_all}. For each line we report respectively: 
wavelenght, $\log \tau_0$, derived frequecy of the principal radial mode (in parentheses the estimated error), $\gamma_0$, semi-amplitude and phase of the variation.
Columns from seventh to tenth reported the parameters derived for the fits of EWs, that is:  relative amplitude, average, semi-amplitude and phase of the 
variations.}
\label{res_fit}     
\centering          
\begin{tabular}{l l | c c c c | c r c c} 
\hline\hline               
                   &                &\multicolumn{4}{c|}{HRV fit parameters}    &\multicolumn{4}{c}{EW fit parameters}                  \\  
\hline
 ~~~~~~~~~~$\lambda$   & $\log \tau_0$  &  freq  & $\gamma_0$     &       K      & $\phi$ & rel. & $\gamma_0$~~~~~~  &    K     &$\phi$  \\
 ~~~~~~~~({\AA})       &                &  (c/d) & (km s$^{-1}$)  & (km s$^{-1}$)&        & (\%)  &  (m{\AA})~~~~     & (m{\AA}) &      \\
\hline                                                                                                                                            
 C{\sc ii}\,6578.052   &  $-$2.11 & 5.24967(1)& $-$3.2\,$\pm$\,0.6 & 15.5\,$\pm$\,0.1 & 0.526\,$\pm$\,0.001 &9.5 &  88.2\,$\pm$\,0.2 & 8.4\,$\pm$\,0.2 & 0.28\,$\pm$\,0.01 \\
 C{\sc ii}\,6582.882   &  $-$1.84 & 5.24967(1)& $-$3.3\,$\pm$\,0.5 & 13.5\,$\pm$\,0.1 & 0.544\,$\pm$\,0.002 &9.9 &  82.4\,$\pm$\,0.2 & 8.1\,$\pm$\,0.2 & 0.19\,$\pm$\,0.01 \\
\hline                                                                                                                                 
 N{\sc ii}\,4607.153   &  $-$0.53 & 5.24969(2)& $-$1.9\,$\pm$\,0.4 & 15.3\,$\pm$\,0.2 & 0.456\,$\pm$\,0.002 &9.1 &  56.0\,$\pm$\,0.2 & 5.2\,$\pm$\,0.3 & 0.19\,$\pm$\,0.01 \\
 N{\sc ii}\,5010.621   &  $-$0.53 & 5.24967(2)& $-$1.8\,$\pm$\,0.5 & 13.1\,$\pm$\,0.2 & 0.530\,$\pm$\,0.002 &7.4 &  51.0\,$\pm$\,0.3 & 3.8\,$\pm$\,0.4 & 0.24\,$\pm$\,0.02 \\
 N{\sc ii}\,5666.629   &  $-$0.97 & 5.24967(2)& $-$3.1\,$\pm$\,0.5 & 14.1\,$\pm$\,0.2 & 0.528\,$\pm$\,0.002 &8.4 &  83.3\,$\pm$\,0.2 & 7.0\,$\pm$\,0.2 & 0.18\,$\pm$\,0.01 \\
 N{\sc ii}\,5676.017   &  $-$0.82 & 5.24967(2)& $-$4.3\,$\pm$\,0.4 & 13.3\,$\pm$\,0.2 & 0.536\,$\pm$\,0.003 &7.9 &  69.4\,$\pm$\,0.2 & 5.5\,$\pm$\,0.2 & 0.19\,$\pm$\,0.01 \\
 N{\sc ii}\,5679.558   &  $-$1.27 & 5.24968(1)& $-$4.5\,$\pm$\,0.6 & 14.3\,$\pm$\,0.2 & 0.500\,$\pm$\,0.002 &7.1 & 120.2\,$\pm$\,0.2 & 8.5\,$\pm$\,0.2 & 0.21\,$\pm$\,0.01 \\
\hline                                                                                                                               
 O{\sc ii}\,4591.010   &  $-$0.82 & 5.24967(1)& $-$5.3\,$\pm$\,0.7 & 15.1\,$\pm$\,0.1 & 0.544\,$\pm$\,0.001 &1.2 & 120.5\,$\pm$\,0.3 & 1.5\,$\pm$\,0.4 & 0.05\,$\pm$\,0.04 \\
 O{\sc ii}\,4661.643   &  $-$0.97 & 5.24966(1)& $-$4.7\,$\pm$\,0.7 & 15.9\,$\pm$\,0.1 & 0.597\,$\pm$\,0.001 &1.9 & 121.5\,$\pm$\,0.3 & 2.4\,$\pm$\,0.4 & 0.39\,$\pm$\,0.03 \\
\hline                                                                                                                                  
Si{\sc iii}\,4552.622  &  $-$1.70 & 5.24967(1)& $-$1.9\,$\pm$\,0.8 & 14.4\,$\pm$\,0.1 & 0.535\,$\pm$\,0.002 &2.5 & 209.0\,$\pm$\,0.2 & 5.2\,$\pm$\,0.3 & 0.33\,$\pm$\,0.01 \\
Si{\sc iii}\,4567.840  &  $-$1.41 & 5.24967(1)& $-$2.6\,$\pm$\,0.7 & 15.0\,$\pm$\,0.1 & 0.556\,$\pm$\,0.001 &2.7 & 174.3\,$\pm$\,0.2 & 4.8\,$\pm$\,0.3 & 0.36\,$\pm$\,0.01 \\
Si{\sc iii}\,4574.757  &  $-$0.97 & 5.24968(1)& $-$3.2\,$\pm$\,0.6 & 15.4\,$\pm$\,0.1 & 0.505\,$\pm$\,0.001 &2.9 & 117.9\,$\pm$\,0.2 & 3.4\,$\pm$\,0.3 & 0.27\,$\pm$\,0.02 \\
\hline               
All                    &          & 5.249677(7) & $-$3.5\,$\pm$\,0.4 & 15.5\,$\pm$\,0.1 & 0.516\,$\pm$\,0.001 &            &               &             &                 \\
\hline 

\end{tabular}
\end{table*}

  \begin{figure}
   \centering
   \includegraphics[width=9cm]{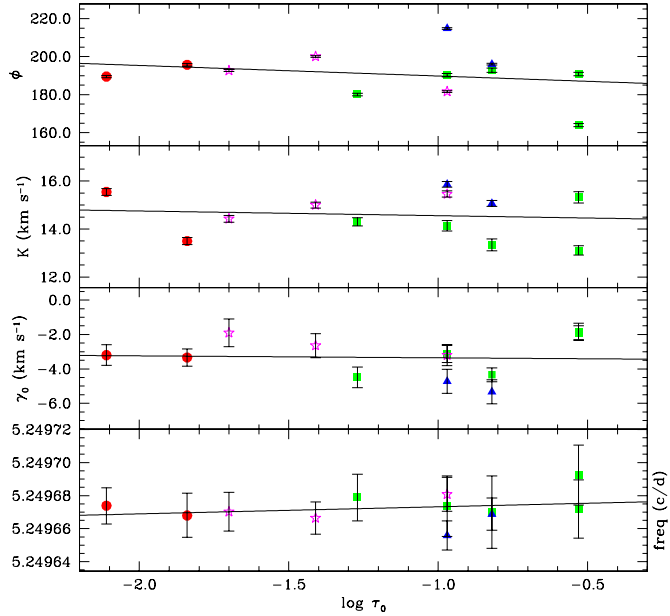}
   \caption{In this figure we report as a function of optical depth the computed frequencies, 
            center of mass ($\gamma_0$), amplitudes and phases derived from the fit of velocities 
            for each of the selected lines. Meaning of the symbols is: circles (red)
            carbon lines, stars (magenta) silicon lines, triangles (blues) oxygen lines
            and boxes (green) nitrogen lines.}
   \label{strat}
   \end{figure}

\section{Data Analisys}
One of the most used code to analyse a large set of periodic data in order to obtain
frequency, velocity of center of mass ($\gamma_0$), amplitude of the variation
(K) and phase ($\phi$), is {\it Period04} developed by \citet{lenz05}. We used this
code to perform a line-by-line sinusoidal fit of our velocities. For each selected line,
we reported in Tab.~\ref{res_fit} frequency,  $\gamma_0$,  amplitude K and phase
$\phi$ and relative errors as determinated by using that code. Each box on the left side 
of Fig.~\ref{pdm} shows the velocities derived line by line folded with the corresponding 
frequency and with the sinusoidal fit over-imposed.

With the aim to search for vertical stratification, we plotted all the quantities reported
in Tab.~\ref{res_fit} regarding the velocity versus optical depths. To compute the optical
depths, we used the code XLINOP \citep{kur81} applied to LTE atmospheric model with
T$_{\rm eff}$\,=\,24000 K, $\log g$\,=\,3,9, solar metalicity and $\xi$\,=\,8 km $^{-1}$.
These atmospheric values have been derived by us in the framework of a detailed abundances
revision that it is currently under study.
As we shown in Fig.~\ref{strat}, there is no evidence of some kind of stratification
through the stellar atmosphere, at least in the range of optical depth probed by our lines.

Since the results obtained line-by-line are consistent, with the aim to improve the
quality of our measurements, we computed the radial velocity 
as an average over the values obtained from all lines, without paying any attention to 
the different chemical species. In this case, the fitting 
procedure described before gave us the results presented in the last row of 
Tab.~\ref{res_fit} and then we adopted it as frequencies of the radial mode of
$\beta$ Cep: $f_1$\,=\,5.249677\,$\pm$\,0.000007~c~d$^{-1}$. The corresponding phase 
diagram is shown in the right side of Fig.~\ref{pdm}. The epoch used for the plot
is that of the minimum radius, that is HJD$_{\rm (R_*\,min)}$\,=\,24\,53579.2649.

Since the study by \citet{pigu92}, it is well known
that $\beta$ Cephei shows apparent changes of its pulsation period
because the light\,-\,time effect induced by the orbital motion of the
binary system. In our conclusion we have neglected this effect since
our time baseline covers scarcely 1.4~$\%$ of the orbital period,
estimated by those authors in 91.6 yrs.

\begin{figure*}
   \centering
   \hbox{
   \includegraphics[width=7cm]{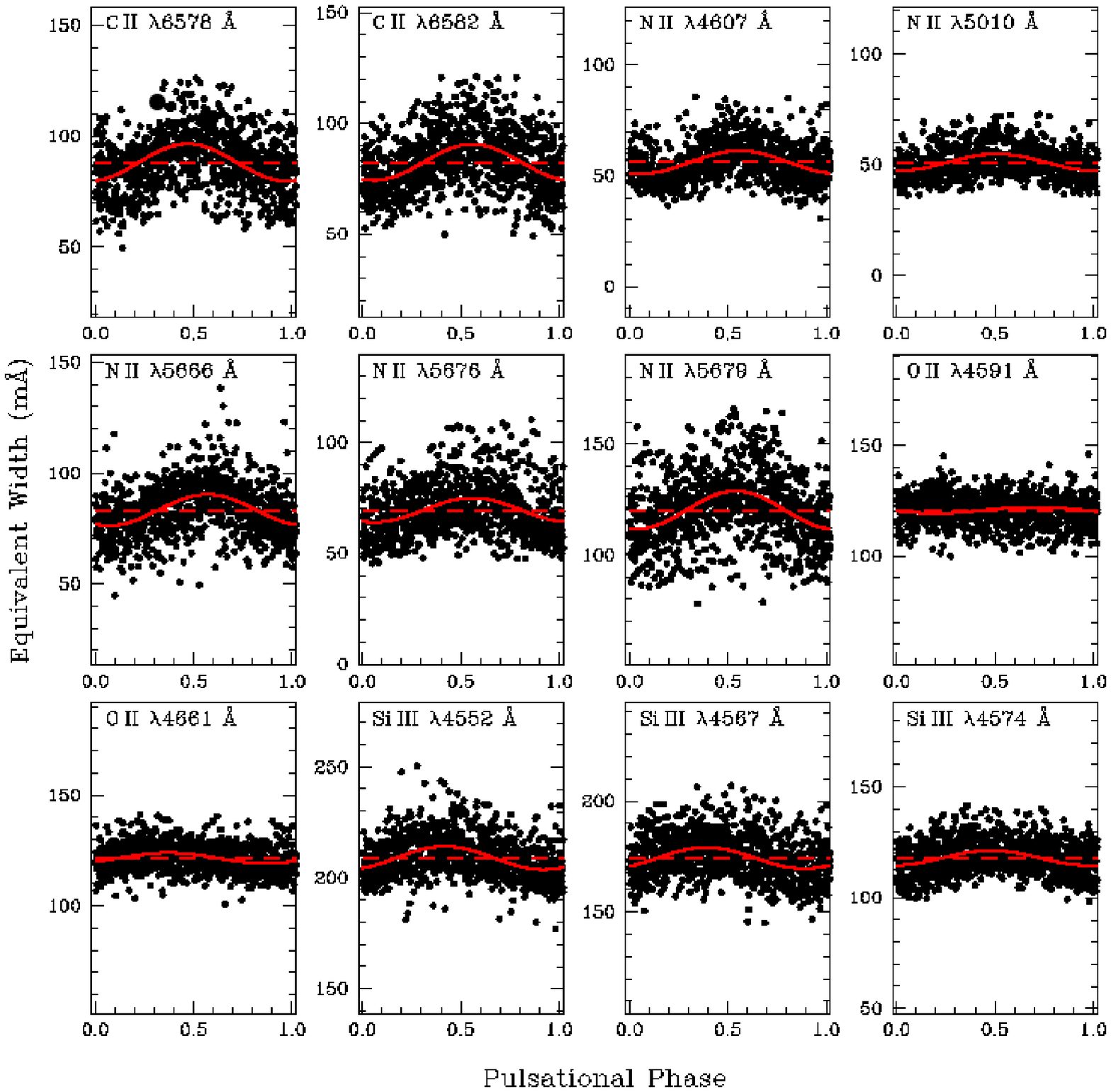}
   \includegraphics[width=7cm]{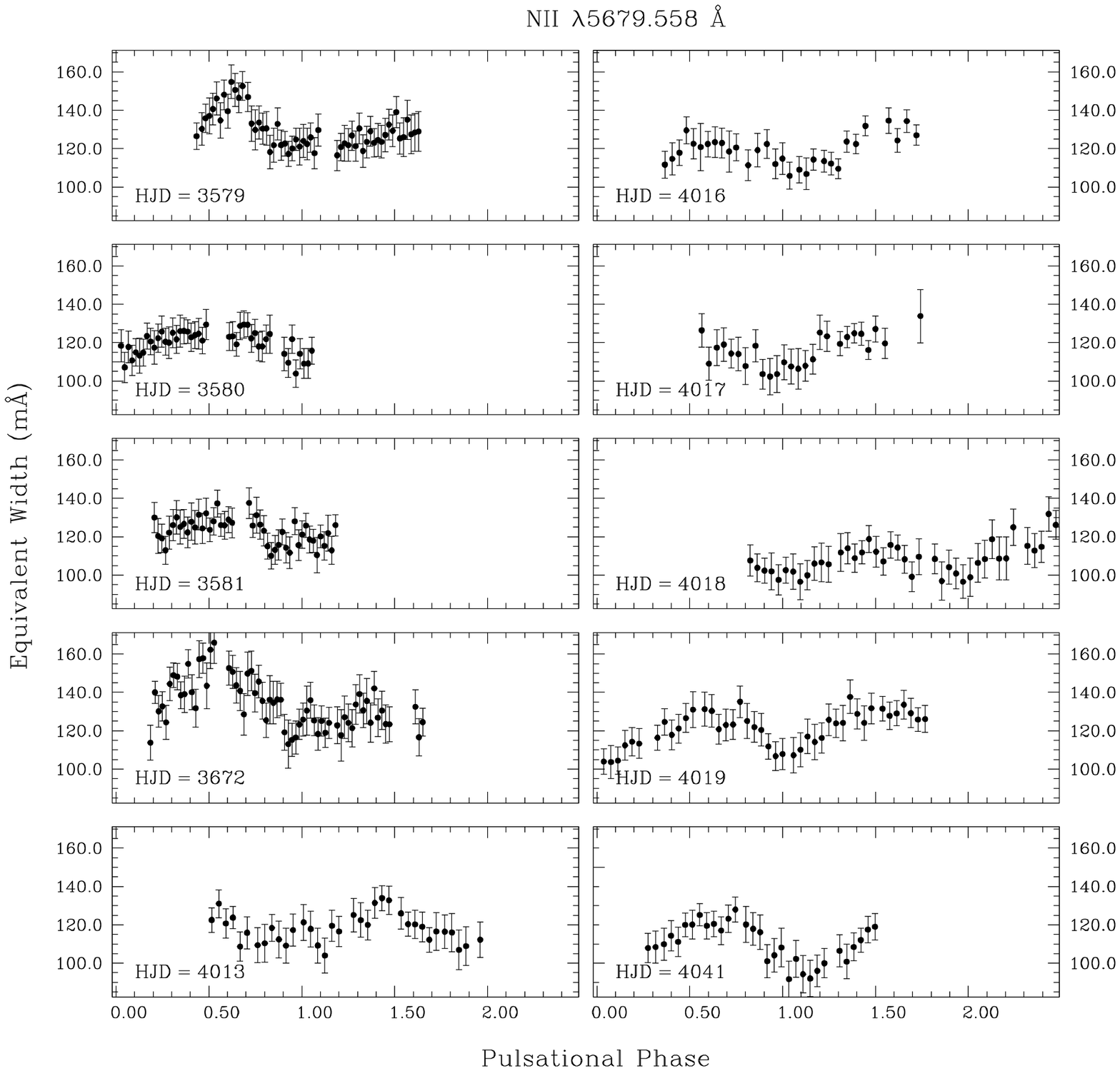}
       }
   \caption{({\it Left side)} Variability of the measured equivalent widths for 
            all the selected lines. Parameters of the sinusoidal fits over-imposed are those 
            reported in Tab.~\ref{res_fit}. ({\it Right side)} Equivalent width 
            variations for the N{\sc ii} $\lambda$5679 {\AA} observed in ten different nights. 
            In each box we indicated the julian day~-~24\,50000.}
   \label{var_all}
   \end{figure*}

As a by-product, we measured also the equivalent widths of each selected line.
In Fig.~\ref{var_all} (left side), we show the phase diagrams obtained for all the 
EW measured for the selected lines phased with the ephemeris derived from the
velocities. As we stated in the previous section, all these lines 
are unblended and sufficiently isolated to avoid contamination from nearby lines.  
Sinusoidal fits over-imposed on the experimental data have been
computed neglecting those points for which their relative errors exceed 20\,$\%$ and
keeping all the points within $\pm$~3\,$\sigma$ from the mean value, where $\sigma$ 
is the standard deviation. The parameters of the sinusoidal fits with
their errors are presented in Tab.~\ref{res_fit}. We did find variations with small
amplitude for some lines as for example O{\sc ii}\,4661~{\AA} and Si{\sc iii} lines 
(relative amplitudes $\le$\,3\,$\%$). A more marked variability, up to $\approx$~10~$\%$, 
has instead been inferred for N{\sc ii} and C{\sc ii} lines.

Despite difficulties in measuring the EW of a spectral line, for instance the
choice of line limits and continuum, the phase diagrams showed a scatter not justified 
if related to errors. For the sake of clarity, we showed in Fig.~\ref{var_all} (right 
side) the EW measured in 10 different nights for the N{\sc ii}~$\lambda$5679~{\AA} line. 
Similar behavior have been observed also in the variations of the others spectral lines 
here considered. A so complex changing, observed night by night, could be ascribed
to the multi-periodic nature of $\beta$ Cep (\citet{aerts94}, \citet{telting97}).
\section{Conclusions and discussion}
Spectroscopic observations of $\beta$ Cephei have been carried out at the stellar 
station of the {\it INAF\,-\,Osservatorio Astrofisico di Catania}, 
in total we collected 932 spectra with a time baseline covering more than one year. 

As for the radial velocities we checked the behavior of velocities coming from different 
lines and for each group of velocities we computed the pulsation frequency of the
radial mode, the $\gamma_0$, amplitude and phase of the relative sinusoidal fit. Plotting
these values as a function of the optical depth of each line we do not see any evidence
of stratification through the stellar atmosphere. For what 
that concerns equivalent widths, we showed their behavior in time and, in the case of the
N{\sc ii} $\lambda$ 5679.558 {\AA}, the possible influence of the non-radial pulsation on 
the amplitude and shape observed night by night. Further, we did find clear variability
for C{\sc ii} and N{\sc ii} lines, while Si{\sc iii} and O{\sc ii} $\lambda$\,4661~{\AA}
show very small amplitude. For O{\sc ii} $\lambda$\,4591~{\AA} we concluded that its
EW is constant with time, at least at the resolving power of our spectra.

At the end, we think that such large data set of data, radial velocities and equivalent widths,
could be useful for a number of studies, like for instance the study of the orbit utilizing the 
light-time effect or frequencies analysis, so we would like put them at astronomical
communinity's disposal.




\begin{thebibliography}{}

\bibitem[Aerts \& De Cat(2003)]{aerts03} Aerts, C., \& De Cat, P. 2003, Space Science Reviews, 105, 453

\bibitem[Aerts et al.(1994)]{aerts94} Aerts, C., Mathias, P., Gillet, D., \& Waelkens, C. 1994, A\&A, 286, 109

\bibitem[Baldry et al.(1998)]{baldry98} Baldry, I. K., Bedding, T. R., Viskum, M., Kjeldsen, H., \& Frandsen, S.
                                        1998, MNRAS, 295, 33

\bibitem[De Ridder et al.(2002)]{deridder02} 
De Ridder, J., Dupret, M.-A., Neuforge, C., \& Aerts, C. 2002, A\&A, 385, 572

\bibitem[Gezari et al.(1972)]{gezari72} Gezari, D. Y., Labeyrie, A., \& Stachnik, R. V. 1972, ApJ, 173, L1

\bibitem[Hadrava \& Harmanec(1996)]{hadrava96} Hadrava, P., \& Harmanec, P. 1996, A\&A, 315, L401

\bibitem[Hartkopf et al.(2001)]{hartkopf01}  
     Hartkopf, W., Mason, B., Wycoff, G., \& McAlister, H. 2001, 
     Fourth Catalog of Interferometric Measurements of Binary Stars

\bibitem[Kurucz \& Avrett(1981)]{kur81} Kurucz, R. L., \& Avrett, E. H. 1981 SAO Special Report, 391

\bibitem[Lenz \& Breguer(2005)]{lenz05} Lenz P., \& Breger M. 2005, CoAst, 146, 53

\bibitem[Morel et al.(2006)]{morel06} 
Morel, T., Butler, K., Aerts, C., Neiner, C., \& Briquet, M. 2006, A\&A, 457, 651

\bibitem[Osaki(1971)]{osaki71} Osaki, Y. 1971, PASJ, 23, 485 

\bibitem[Pigulsky \& Boratyn(1992)]{pigu92} Pigulsky, A. \& Boratyn, D. A. 1992, A\&A, 253, 178

\bibitem[Schnerr et al.(2006)]{schnerr06} 
  Schnerr, R. S., Henrichs, H. F., Oudmaijer, R. D., \& Telting J. H. 2006, A\&A, 459, L21

\bibitem[Stellingwerf(1978)]{stell78} Stellingwerf, R. F. 1978, ApJ, 224, 953

\bibitem[Sterken \& Jerzykiewicz(1992)]{sterken92} Sterken, C., \& Jerzykiewicz, M. 1992, Space Science Reviews, 62, 95

\bibitem[Telting et al.(1997)]{telting97} Telting, J. H., Aerts, C., \&  Mathias, P. 1997, A\&A, 322, 493

\bibitem[Udry et al.(1999)]{udry99} Udry, S., Mayor, M., \& Queloz, D. 1999, Precise Stellar Radial 
                    Velocities, IAU Colloq. 170, ed. J. B. Heamshaw, \& C. D. Scarfe,
                    ASP Conf. Ser., 185, 367


\end{thebibliography}
\end{document}